\documentclass[conference]{IEEEtran}
\IEEEoverridecommandlockouts
\usepackage{cite}
\usepackage{amsmath,amssymb,amsfonts}
\usepackage{algorithmic}
\usepackage{graphicx}
\usepackage{textcomp}
\usepackage{xcolor}
\usepackage{hyperref}
\begin{document}

\newcommand{\ds}{\textsc{DeepSec}}

\title{A critique of the \ds{} \emph{Platform for \\ Security Analysis of Deep Learning Models}\vspace{-.2em}}

\author{\IEEEauthorblockN{Nicholas Carlini, \textit{Google Research}\vspace{-.2em}}}

\maketitle

\begin{abstract}
At IEEE S\&P 2019, the paper \emph{``\ds{}: A Uniform Platform for
  Security Analysis of Deep Learning Model''} aims to to ``systematically evaluate
the existing adversarial attack and defense methods.'' While the paper's goals
are laudable, it fails to achieve them and presents results that are fundamentally
flawed and misleading. We explain the flaws in the \ds{} work,
along with how its analysis fails to meaningfully evaluate the various attacks
and defenses. Specifically, \ds{} (1) evaluates each defense obliviously,
using attacks crafted against undefended models; (2) evaluates attacks and defenses
using incorrect implementations that greatly under-estimate their effectiveness;
(3) evaluates the robustness of each defense as an average, not
based on the most effective attack against that defense; (4) performs several
statistical analyses incorrectly and fails to report variance; and, (5) as a result
of these errors draws invalid conclusions and makes sweeping generalizations.
\end{abstract}


\section{Introduction}
\ds{} is a platform to ``measure the vulnerability of
[deep learning] models'' and ``conduct  comparative  studies
on  attacks/defenses'' \cite{deepsec}.
Ling \emph{et al.} re-implemented many common attacks
and defenses  to fit the consistent interface of \ds{}
and then
``systematically evaluate the existing adversarial attack and defense methods''
\cite{deepsec}.

Unfortunately, \ds{}'s attack and defense re-implementations,
experimental design, and analysis results are fundamentally
flawed such that the obtained results greatly misrepresent
the effectiveness of both attacks and defenses.
Here, we summarize the ways in which the \ds{} paper errs in its implementation
and analysis.

\section{Summary of Errors}

\vspace{.5em}
\noindent \mbox{\textbf{Evaluation uses attacks that are oblivious to defenses.}}
A security defense can only be meaningfully evaluated by measuring the
effectiveness of attacks crafted against it.
For example, RC2-aware attacks can assess the strength of RC2 encryption,
but using the exact linear characteristic that breaks
DES to assess the strength of RC2 would be meaningless.
In \ds{}, the effectiveness of defenses is measured by crafting attacks on an
undefended model and replaying those attacks on the defended models,
which undermines the stated purpose of its security evaluation.

It might be of some interest whether attacks transfer from undefended to
defended models, however this is not what the \ds{} paper claims to present.
The paper claims to measure ``non-adaptive and white-box'' robustness of defenses;
such ``white-box'' attack evaluations \emph{must} still be run (possibly unmodified) given direct access to each defense \cite{carlini2019evaluating}.
In short, \ds{} in effect performs a \emph{non-adaptive, black-box, 
  zero-query, and transfer-only} attack analysis.

\vspace{.5em}
\noindent \textbf{Defenses are evaluated by average (not minimum) efficacy.}
A key factor differentiating security (and adversarial
robustness) from general forms of fault tolerance is the requirement for worst-case
analysis. Instead, \ds{} uses averages to assess the effectiveness
of different attacks and defenses.

For example, in Table V, \ds{} bolds the column for the \emph{NAT} defense when evaluated on
CIFAR-10 because it gives the highest ``average security'' against all
attacks. However, this is fundamentally the wrong evaluation: actual attacks
have found that NAT has a strictly
lower accuracy \cite{athalye2018obfuscated} under all distortion metrics than
the alternate approach of Madry et al. \cite{madry2018towards}.

\vspace{.5em}
\noindent \textbf{Multiple attacks are implemented incorrectly.}
Table XIV in \ds{} reports an attack misclassification rate substantially lower
than in prior work. For example, on MNIST with a $\ell_\infty$ distortion bound of
$\varepsilon=0.3$, \ds{} reports the attack success rate of FGSM
\cite{goodfellow2014explaining} is 30.4\%.
Our re-implementation of FGSM reaches a 66\% success rate on their model\cite{githubissue}.

\ds{} also reports a 76\% success rate with JSMA \cite{papernot2016limitations}
for $\gamma=0.1$. However, this attack reaches a 95\% success rate when using
the official implementation in CleverHans \cite{papernot2016technical}.
Further, the reported attack success rates at
$\varepsilon=0.3$ for BIM \cite{kurakin2016adversarial} is 75.6\% and PGD is 82.4\%, contradicting the
100\% success rate reported in the relevant prior work \cite{madry2018towards}.

\vspace{.5em}
\noindent \textbf{The PGD defense is implemented incorrectly.}
While its underlying idea is simple---repeatedly generate and
train on adversarial examples---PGD adversarial training (PAT)
is very difficult to get right in practice.
The authors claim to evaluate the approach of Madry et al. \cite{madry2018towards}
but make at least three errors:
\begin{itemize}
\item \emph{Incorrect loss function.} PAT should train only on adversarial
  examples, but \ds{} also uses clean data.

\item \emph{Incorrect model architectures.} PAT specifies large model capacity
  is required, but \ds{} uses a small model.

\item \emph{Incorrect hyperparameter settings.} PAT should train for 83 epochs
  to converge, but \ds{} trains for only 20.
\end{itemize}
Possibly because of these implementation differences, \ds{} incorrectly concludes
that a weak form of adversarial training \cite{goodfellow2014explaining}
performs better than PGD adversarial training, contradicting prior results 
\cite{athalye2018obfuscated,madry2018towards}.

\vspace{.5em}
\noindent \textbf{No error bars for any results.}
The \ds{} paper does not include any information about the variance of its
analysis results.
When we run the authors' FGSM attack implementation 16 times we observe
an attack success rate that is approximately normal with a mean of 32.7\% and
standard deviation of 6.8\%. Such high variance would make many pair-wise
comparisons in Table XIV not be statistically significant.

\vspace{.5em}
\noindent \textbf{Analysis computes averages over different threat models.}
The \ds{} report computes the mean over different threat models, which gives a
number that is completely uninformative. When \ds{} reports 60\% robustness for
a defense, this means the following:
first, the adversary chooses a random threat model with a certain probability
($\ell_\infty$: 50\%, $\ell_0$:  5\%, and $\ell_2$: 45\% of the time);
then, the adversary chooses a random attack from those studied with that threat model;
then, for that chosen attack, the attacker will fail 60\% of the time.
Of course, no attacker would follow this protocol,
and therefore, this this across-threat-model average is not informative.

\vspace{.5em}
\noindent \textbf{The permitted distortion ($\varepsilon$) is too large to be meaningful.}
The purpose of an $\ell_p$ distortion bound is to ensure the true label can not change
\cite{goodfellow2014explaining, carlini2019evaluating}.
The \ds{} paper studies a CIFAR-10 $\ell_\infty$ distortion of $\varepsilon=0.1$ and
$\varepsilon=0.2$, which is $3\times$ (or $6\times$)
larger than what is typically studied \cite{athalye2018obfuscated,madry2018towards}.

The paper further studies $\ell_\infty$ distortion bounds as high as $0.6$
in Table VII, and $0.5$ in Table XIV and XV.
These extreme distortions would allow any image to be converted
to solid grey, and (for $0.6$) past that to a low-contrast version of any other image,
contradicting the purpose of $\ell_p$ threat models.

\vspace{.5em}
\noindent \textbf{Attacks and defenses evaluated ignoring threat model.}
Both attacks and defenses are typically designed to target a specific set
of threat models. All of the attacks considered were designed to minimize
exactly one specific distortion metric; \ds{}, however, evaluates all attacks
on every metric without optimize the attack for each metric.

Even more concerning, the majority of the defenses studied contain explicit
threat models explicitly scoping their contributions to limited attack
models
(e.g., PAT is only designed to be robust to $\ell_\infty$ attacks with $\varepsilon<0.031$ on CIFAR-10 \cite{madry2018towards}).
\ds{} performs unfair defense evaluations
by violating the threat model of every defense which contains one (e.g.,
by evaluating PAT against $\ell_0$ and $\ell_1$ attacks).
When defenses are evaluated under different threat models than
originally stated, this fact should be stated explicitly.

\vspace{.5em}
\noindent \textbf{Incorrect experimental design for comparing attacks.}
The numbers presented in Table V do not make it possible to compare how
well different attacks perform on defended models.
While ILLC is a much stronger attack than LLC (as shown in the original paper
\cite{kurakin2016adversarial}),
the \ds{} report makes it appear that LLC is a
better attack against defended models.
This is due to flawed experimental design: \ds{} does not evaluate defenses on all the relevant examples,
but only on those that fool the baseline model.
Therefore, the $39\%$ attack success rate of LLC against PAT is computed from
only 134 of the 1000 possible attack samples; in contrast, ILLC's
$16.3\%$ success rate is computed from \emph{all} of the 1000 samples.
These numbers are fundamentally incomparable.

\vspace{.5em}
\noindent \textbf{Anomalies due to incorrect experimental design.}
When given strictly more power, the adversary
should never do worse. However, Table VII reports that an MNIST attacker
is \emph{less} likely to succeed with large permitted $\ell_\infty$ perturbation
of at most 0.6 compared to the \emph{smaller} budget of at most 0.2.




\noindent
\textbf{Sweeping and false conclusions.}
In multiple places, the \ds{} paper states all defenses are ``more or less''
effective \cite{deepsec}, which is false. 
Most of the defenses studied offer 0\% robustness to any of the currently-known
state-of-the-art white-box or black-box attacks \cite{athalye2018obfuscated}.
Instead, all of the paper's conclusions should be restricted to
\emph{non-adaptive, black-box, zero-query, and transfer-only} adversarial examples.

\section{Conclusion}

Improperly-performed experiments are worse than experiments not performed
when published as authoritative results.
Because survey papers have significant influence on the
understanding of
the academic community,
researchers that craft such papers should take great care to ensure the accuracy of all their
results and not introduce misinformation.
Unfortunately, the analysis of \ds{} \cite{deepsec}
falls below this
bar due to fundamental flaws in its experimental design
and evaluation.

Researchers who set out to reproduce prior work
must hold themselves
to an exceptionally high standard.
Of the 4 attacks and 1 defense implementations in \ds{} that we studied,
all had at least one significant flaw.
Clean-room re-implementations can be extremely valuable to
ensure correctness of reported results;
however, after reproducing prior work, it is critical to compare
to existing implementations.
For all of these attacks and defenses, correct and open-source implementations
already exist in CleverHans \cite{papernot2016technical} but these
were not used or compared against by the \ds{} authors.

Future work should not follow the evaluation approach
taken by the \ds{} paper.
The \ds{} framework itself should not be used to evaluate defenses until
all remaining attacks and defenses are confirmed to be correct.
The analysis results of Tables V, VI, and VII should be disregarded
except insofar as they analyze the transferability of adversarial examples.
The sweeping general conclusions should be ignored.

We refer the interested reader to \cite{carlini2019evaluating}
for a longer discussion of common flaws, and recommendations for how they can be
best avoided, when evaluating adversarial robustness.



\bibliographystyle{IEEEtranS}
\bibliography{paper}

\begin{thebibliography}{1}
\providecommand{\url}[1]{#1}
\csname url@samestyle\endcsname
\providecommand{\newblock}{\relax}
\providecommand{\bibinfo}[2]{#2}
\providecommand{\BIBentrySTDinterwordspacing}{\spaceskip=0pt\relax}
\providecommand{\BIBentryALTinterwordstretchfactor}{4}
\providecommand{\BIBentryALTinterwordspacing}{\spaceskip=\fontdimen2\font plus
\BIBentryALTinterwordstretchfactor\fontdimen3\font minus
  \fontdimen4\font\relax}
\providecommand{\BIBforeignlanguage}[2]{{%
\expandafter\ifx\csname l@#1\endcsname\relax
\typeout{** WARNING: IEEEtranS.bst: No hyphenation pattern has been}%
\typeout{** loaded for the language `#1'. Using the pattern for}%
\typeout{** the default language instead.}%
\else
\language=\csname l@#1\endcsname
\fi
#2}}
\providecommand{\BIBdecl}{\relax}
\BIBdecl

\bibitem{athalye2018obfuscated}
A.~Athalye, N.~Carlini, and D.~Wagner, ``Obfuscated gradients give a false
  sense of security: Circumventing defenses to adversarial examples,''
  \emph{arXiv preprint arXiv:1802.00420}, 2018.

\bibitem{githubissue}
N.~Carlini, https://github.com/kleincup/DEEPSEC/issues/3, 2019.

\bibitem{carlini2019evaluating}
N.~Carlini, A.~Athalye, N.~Papernot, W.~Brendel, J.~Rauber, D.~Tsipras,
  I.~Goodfellow, and A.~Madry, ``On evaluating adversarial robustness,''
  \emph{arXiv preprint arXiv:1902.06705}, 2019.

\bibitem{goodfellow2014explaining}
I.~J. Goodfellow, J.~Shlens, and C.~Szegedy, ``Explaining and harnessing
  adversarial examples,'' \emph{arXiv preprint arXiv:1412.6572}, 2014.

\bibitem{kurakin2016adversarial}
A.~Kurakin, I.~Goodfellow, and S.~Bengio, ``Adversarial examples in the
  physical world,'' in \emph{ICLR (Workshop Track)}, 2016.

\bibitem{deepsec}
X.~Ling, S.~Ji, J.~Zou, J.~Wang, C.~Wu, B.~Li, and T.~Wang, ``Deepsec: A
  uniform platform for security analysis of deep learning model,'' in
  \emph{IEEE Symposium on Security and Privacy}, 2019.

\bibitem{madry2018towards}
A.~Madry, A.~Makelov, L.~Schmidt, D.~Tsipras, and A.~Vladu, ``Towards deep
  learning models resistant to adversarial attacks,'' \emph{ICLR}, 2018.

\bibitem{papernot2016technical}
N.~Papernot, F.~Faghri, N.~Carlini, I.~Goodfellow, R.~Feinman, A.~Kurakin,
  C.~Xie, Y.~Sharma \emph{et~al.}, ``Technical report on the cleverhans v2. 1.0
  adversarial examples library,'' \emph{arXiv:1610.00768}, 2016.

\bibitem{papernot2016limitations}
N.~Papernot, P.~McDaniel, S.~Jha, M.~Fredrikson, Z.~B. Celik, and A.~Swami,
  ``The limitations of deep learning in adversarial settings,'' in
  \emph{EuroS\&P}, 2016.

\end{thebibliography}

\end{document}